\documentclass[twocolumn,english]{revtex4-1}
\usepackage[latin9]{inputenc}
\usepackage{verbatim}
\usepackage{amstext}
\usepackage{amssymb}
\usepackage{cancel}
\usepackage{graphicx}

\makeatletter
\@ifundefined{date}{}{\date{}}
\makeatother

\usepackage{babel}
\begin{document}
\global\long\def\ket#1{\left|#1\right\rangle }

\global\long\def\bra#1{\left\langle #1\right|}

\global\long\def\braket#1#2{\left\langle #1\left|#2\right.\right\rangle }

\global\long\def\ketbra#1#2{\left|#1\right\rangle \left\langle #2\right|}

\global\long\def\braOket#1#2#3{\left\langle #1\left|#2\right|#3\right\rangle }

\global\long\def\mc#1{\mathcal{#1}}

\global\long\def\nrm#1{\left\Vert #1\right\Vert }

\title{Collective operation of quantum heat machines via coherence recycling,
and coherence induced reversibility}

\author{Raam Uzdin}

\email{raam@mail.huji.ac.il}

\selectlanguage{english}%

\affiliation{Fritz Haber Research Center for Molecular Dynamics, Hebrew University
of Jerusalem, Jerusalem 9190401, Israel\\
Department of Chemistry and Biochemistry, \\
University of Maryland, College Park, Maryland 20742, USA}
\begin{abstract}
Collective behavior where a set of elements interact and generate
effects that are beyond the reach of the individual non interacting
elements, are always of great interest in physics. Quantum collective
effects that have no classical analogue are even more intriguing.
In this work we show how to construct collective quantum heat machines
and explore their performance boosts with respect to regular machines.
Without interactions between the machines the individual units operate
in a stochastic, non-quantum manner. The construction of the collective
machine becomes possible by introducing two simple quantum operations:
coherence extraction and coherence injection. Together these operations
can harvest coherence from one engine and use it to boost the performance
of a slightly different engine. For weakly driven engines we show
that the collective work output scales quadratically with the number
of engines rather than linearly. Eventually, the boost saturates and
the scaling becomes linear. Nevertheless, even in saturation, work
is still significantly boosted compared to individual operation. To
study the reversibility of the collective machine we introduce the
'entropy pollution' measure. It is shown that there is a regime where
the collective machine is $N$ times more reversible while producing
$N$ times more work, compared to the individual operation of $N$
units. Moreover, the collective machine can even be more reversible
than the most reversible unit in the collective. This high level of
reversibility becomes possible due to a special symbiotic mechanism
between engine pairs.

\end{abstract}
\maketitle

\section{Introduction }

When particles or systems interact in a special synchronized way,
they can collectively accomplish remarkable tasks that are far beyond
the capabilities of the sum of individual units. An electrons Cooper
pair can flow in a solid without resistance while unpaired electrons
cannot. In biology, symbiotic relationship refers to the case where
two different species benefit from interacting with each other. Similarly,
in this work we study the collective operation of quantum heat machines.
The individual units are functional heat machines. The collective
machine that emerges by introducing a special synchronized interaction,
is also a heat machine but with superior performance. Furthermore,
the synchronized interaction is based on coherence exchange between
the units and is therefore inherently quantum. Since this interaction
affects only coherences in the energy basis, the machines do no exchange
heat or work with each other. In a certain case we show that the machines
can be divided to two species that exhibit symbiotic interrelation.

A heat engine is a device that converts heat flows into useful work.
Work may take different forms. In the quantum regime it may be, for
example, the amplification of light (lasing medium). Alternatively
a quantum power refrigerator may use microwave radiation to take away
heat from a device and dispose it in some ambient environment. On
top of the standard uses of heat machines, it has been suggested to
use quantum heat machines for non classical tasks such as entanglement
creation \cite{HuberEntangGenHeat} and state squeezing \cite{NoriHeatSqueezing}. 

With the growing technological abilities to control and manipulate
single atoms and ions, the construction of microscopic quantum heat
machines is within reach. A single atom heat engine and an NMR refrigerator
have already been built \cite{rossnagelIonEngExp,AlgoCool2005expNature}.
Suggestions for realizations in several other quantum systems include
quantum dots \cite{QuantDotEngine,BergenfeldtQDotsMicrowave}, superconducting
devices \cite{PekolaSCengine,HuberEntangGenHeat,campisi2014FT_SolidStateExp,pekola2015towards},
cold bosons \cite{FialkoColdBosonsEng}, and optomechanical systems
\cite{mari2012,ZhangOptoMechEng,OptNanomechEng}. In particular, the
rapid advancement in fabrication technology of superconducting circuits
may open the road to on chip heat machines.

The study of quantum heat machines started in \cite{scovil59} and
has become an increasingly active area of research in recent years
(see \cite{alicki79,k24,k152,k221,levy14,rahav12,allmahler10,linden10,mahler07b,skrzypczyk2014work,gelbwaser13,kolar13,alicki2014quantum,Nori2007QHE,lutz14,BinderOperational,Correa2014EnhancedSciRep,dorner2013extracting,palao13,dorner2012emergent,DelCampo2014moreBang,malabarba2014clock,Gelbwaser2015Rev,WhitneyThermElec,AllahverdyanOptDualStroke,correa2015InternalLeaks,mari2012,SeifertCohFeedbackEng,AlgoCoolQuantGas2011Nature,AlgoCool2005expNature,boykin2002algorithmic,EspositoFastDriven,MartiWorkCorr,mahler07,segal06,schulman2005physical,tal02,Correa2016Feedback,PopescuSallMaxEff}
for a very partial list). One of the main goals in the study of quantum
heat machines is to discover and understand the fundamental differences
between small quantum heat machines and their macroscopic counterparts. 

Surprisingly, quantum heat machines show a striking resemblance to
their classical counterparts and often exhibit classical features.
However, recent studies have isolated several strictly quantum features.
In \cite{Anders2015MeasurementWork} a protocol was presented to extract
work from coherences using a thermal bath without changing the energy
population. In \cite{MitchisonHuber2015CoherenceAssitedCooling} it
was shown that coherent quantum dynamics can lead to faster cooling
rates compared to classical dynamics. In \cite{EquivPRX} it was shown
that the coherence of the working fluid in the energy basis \footnote{In contrast to quantum information, in quantum thermodynamics the
energy basis is a preferred basis. This is due to the fact that quantum
thermodynamics has a preferred set of observables e.g. heat, work
and internal energy. } gives rise to a 'thermodynamic equivalence principle' of the three
main engine types: four-stroke, two-stroke, and continuous engine.
In the equivalence regime these machines still have entirely different
dynamics, yet their thermodynamics features are the same. This result
was recently extended to the non-Markovian regime where the machine
and the bath are strongly coupled \cite{RUnonMarkovianEquiv}. For
other studies of coherence in quantum heat engines see \cite{mukamel12,HarbolaEngCoh4,gelbwaserEnhanDegen,SeifertCohFeedbackEng}.

More relevant to the present work, is the identification of two distinct
work extraction mechanisms \cite{EquivPRX}: a coherent (quantum)
mechanism, and a stochastic (classical) mechanism. The coherent mechanism
operates on coherences in the energy basis while the stochastic mechanism
operates on energy populations (population inversion). Thus to observe
quantum effects in power and heat as in \cite{EquivPRX}, the coherent
mechanism must be more pronounced than the stochastic mechanism. For
this to occur it is critical that the thermal strokes be significantly
shorter than the thermalization time. As a result the work stroke
(unitary evolution) starts with non zero coherences and the coherence
work extraction becomes important. For a study of the role of coherence
in specific machines see \cite{mukamel12,MitchisonHuber2015CoherenceAssitedCooling}.
For additional studies of coherence in quantum thermodynamics see
\cite{Anders2015MeasurementWork,aaberg2014catalytic,malabarba2014clock,LostaglioRudolphCohConstraint,RudolphPRX_15,BinderOperational,KamilCohWork,dorner2013extracting,nieuwenhuizen}. 

One of the finding of the present work is that even if \textit{only}
the stochastic work extraction mechanism operates (i.e, the device
is a stochastic machine), there is still a quantum resource that can
be exploited for other purposes. This quantum resource is coherence.
Although coherence is regenerated in each cycle in the engine by interacting
with the classical field, it does not participate in the stochastic
work production process. Hence, the residual coherence can be used
and consumed for other purposes without affecting the performance
of the engine. It is shown here that \textit{coherence obtained from
one engine can be injected to a second engine, and boost its power
output by activating its coherent work extraction mechanism}. The
present paper studies the possibilities entailed in coherence alteration
processes in quantum heat machines. These processes not only give
rise to new types of devices, but also have profound implications
on reversibility and entropy balance in quantum heat machines. To
illustrate our findings we have chosen the smallest and simplest heat
engine model that can exhibit them. Nevertheless, our findings are
general and can be applied to other more complex devices.

Recently, several studies considered quantum heat engines with a working
fluid that contains multiple particles \cite{campisiUnivPhaseTran,DelCampoManyParticleEng,HardalOzgurSuperrad}.
In \cite{OzgurSyncSpin} a central spin is used to synchronize the
other particles. These very interesting studies should be distinguished
from the present study which focuses on \textit{multiple engines}
that operate collectively.

Section II introduces the basic unit that will be used to construct
a collective machine. In section III the over-thermalization regime
in quantum heat machines is briefly described. Then, in section IV
we introduce the coherence extraction (CE) and coherence injection
(CI) processes. In section V we quantify the amount of entropy reduction
associated with CE. Next in section VI the 'entropy pollution' irreversibility
measure is introduced and used to show the impact of CE on non collective
heat machines. Section VII studies collective machine operation where
CE and CI are used for coherence sharing between the engines. After
studying the performance boost gained by the collective operation
we give our conclusion in section VIII.

\section{The basic unit - a two stroke three level quantum heat machine }

\begin{figure}[h]
\includegraphics[width=8.6cm]{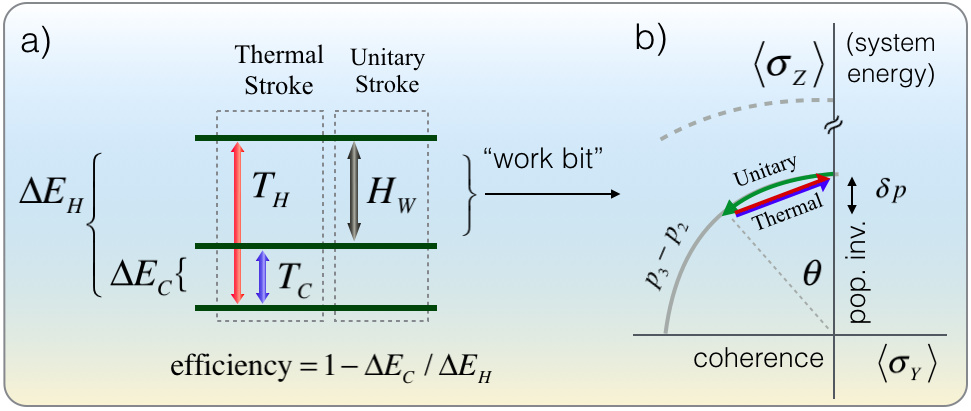}

\caption{(a) The three-level two-stroke engine used as a basic unit for the
collective machine. In the thermal stroke the hot bath operates on
levels 1 \& 3 and the cold bath on levels 1 \& 2. The unitary stroke
that follows reduces the energy of the system and extract work. (b)
The engine dynamics can be visualized on the Bloch sphere of level
2 \& 3. The $\left\langle \sigma_{y}\right\rangle $ coordinate corresponds
to coherence and the change in the $\left\langle \sigma_{z}\right\rangle $
coordinate determines the change in population inversion $\delta p$
and consequently also the change in the energy of the system. The
top dashed arch corresponds to the pure states shell of the Bloch
sphere. The mixed state dynamics of the engine takes place inside
the Bloch sphere. The maximal population inversion is determined by
the previous thermal stroke (red-blue arrow). An constant radius curve
with an angle $\theta$ (green) is formed by the unitary evolution
of the work stroke (generated by $H_{w}$). In the over thermalization
regime the unitary stroke generates coherence and the thermal stroke
fully erases them. Thus, in the above setup there are no initial coherences
in the work stroke and the device operates as a stochastic machine
with no quantum effects. }
\end{figure}
The basic machine unit used in this paper is shown in Fig 1a. It is
a simple two-stroke quantum engine. Strokes are time segments in which
the engine undergoes different transformations \cite{EquivPRX}. In
the thermal stroke levels 1 and 3 are connected to a hot bath, while
levels 1 and 2 are connected to a cold bath. After a period that sufficiently
exceeds the thermalization time, the populations are determined by
the Gibbs factors $p_{2}/p_{1}=e^{-\Delta E_{c}/T_{c}},\:p_{3}/p_{1}=e^{-\Delta E_{h}/T_{h}}$.
In the work stroke that follows, the baths are disconnected and a
unitary operation is applied to levels 2 and 3. Work is extracted
during the unitary operation (e.g. the unitary can be a $\pi$ pulse
that takes the particle to a lower energy state and emits an extra
photon in the process). This is a two-stroke quantum heat engine \cite{AllahverdyanOptDualStroke}.
For more information about various engine types see \cite{EquivPRX}.
For a more general introduction to quantum heat machines see reviews
\cite{k281,goold2015review,SaiJanetReview,millen2015review} and references
therein. These reviews also cover other topics of interest in the
more general and rapidly growing field of quantum thermodynamics.
When the bath are disconnected (work stroke) the Hamiltonian of the
engine shown in Fig. 1 is 
\begin{eqnarray}
H & = & H_{0}+H_{w}(t)=\Delta E_{c}\ketbra 22+\Delta E_{h}\ketbra 33\label{eq: H eng}\\
 &  & +\epsilon(t)\cos[(E_{3}-E_{2})t](\ketbra 23+\ketbra 32).
\end{eqnarray}
Although we will use this model for illustration, our findings are
more general and are not limited to this specific three-level model
Hamiltonian. The interactions with the baths are described by Lindblad
operators that depend on temperature.

The engine in Fig. 1 is an Otto engine. Unlike the Carnot engine,
the thermal stages are isochores and not isotherms. This means that
during the thermal stage the 'volume' is fixed and not the temperature.
In discrete quantum heat machine a constant volume means that the
energy levels are not changed in time (as they must change, in a Carnot
machine). Since presently it seems that isochores are easier to implement
compared to isotherms, Otto engines are the default choice in the
quantum regime. The efficiency of Otto engines with uniform level
compression in the work stroke \cite{RUswap}, and also for the current
engine, is $\eta=\frac{W}{Q_{h}}=1-\Delta E_{c}/\Delta E_{h}$ where
$W$ is the work per cycle and $Q_{h}$ is the heat taken from the
hot bath. We point out that thermodynamic bounds on efficiency refer
to steady state operation and not to transients or single shot non
periodic evolution. Although the efficiency in this case does not
depend on the temperature, it is still limited by the Carnot efficiency
$\eta=1-\Delta E_{c}/\Delta E_{h}\le1-T_{c}/T_{h}=\eta_{c}$. When
$\Delta E_{c}/\Delta E_{h}\ge T_{c}/T_{h}$ the machine starts to
work as a refrigerator rather than as an engine. 

In \cite{scovil59} (see \cite{k102} for a more detailed analysis)
it was shown that the Otto engine-refrigerator crossover point (sometimes
called the Carnot point) takes place when the baths cease to generate
population inversion. It is important to point out that this engine-refrigerator
crossover is valid for ``standard'' quantum heat machines where
only the baths appear in the entropy balance over one cycle. In this
paper we introduce the coherence injection process. This process changes
the entropy balance and therefore enables the machine to perform as
an engine even without population inversion. This is related to lasing
without inversion \cite{ScullyLWI}, but has a more general quantum
thermodynamic explanation based on the notions of active and passive
states as explained later on (see also Sec. III.D of \cite{EquivPRX}). 

In this study the unitary is executed by a classical field (e.g.,
a laser light, an RF pulse, or a slow magnetic field). This is experimentally
and theoretically well motivated in a wide range of physical systems
in atomic physics. Some studies replace the classical field by a quantum
battery (work repository) \cite{alicki13,hovhannisyan13,malabarba2014clock,aaberg2014catalytic,RUnonMarkovianEquiv}.
This is very interesting, and deserves further study. Nevertheless,
the classical field framework is highly useful in many quantum systems
where second quantization effects are not important. The classical
field is in some sense a source of coherence (see, for example, \cite{KamilCohWork}).
However its coherence is so abundant that when  the classical field
approximation is valid there is no need to discuss the degradation
of the classical field \cite{KamilCohWork}. In short, our framework
is no different from that used in NMR or common laser and atomic physics
setups.

\section{Bloch sphere engine dynamics in the over-thermalization regime}

In the over-thermalization regime the thermal baths are connected
for periods significantly longer than the thermalization time, so
for all practical purposes coherences are wiped out. Consequently,
the population dynamics in the unitary stroke can be fully described
by stochastic means (the unitary can be replaced by a doubly stochastic
operator \cite{EquivPRX}). Since the population dynamics in the Markovian
thermal stroke can also be fully described by stochastic means, the
quantum machine in this regime operates as a fully stochastic device.
Thermodynamic observables such as work, heat and baths entropy generation,
depend only on energy populations and not on energy coherences. Thus,
in this regime there is little hope of observing quantum effects unless
the scheme is significantly modified. The 'over' prefix in the term
over-thermalization \cite{EquivPRX} is due to the fact that beyond
a certain point, large thermalization rate degrades the performance
of the engine, as it mitigates the coherent work extraction mechanism.
This degradation due to over-thermalization has been previously observed
in specific models in \cite{levy14,k102}, but in \cite{EquivPRX}
it was understood as a generic effect. In this work we show that even
in this seemingly classical-stochastic regime of over-thermalization,
quantum thermodynamic effects can be observed. 

For work extraction, the exact details of the thermalization mechanism
are not important since throughout the paper we assume over-thermalization.
However for bath entropy generation accounting we do use the weak
coupling result to relate the heat and the bath entropy change via
$dQ=-TdS_{bath}$ where $T$ is the temperature of the bath. The weak
coupling limit is widely applicable in many physical systems and constitutes
the standard approach in quantum open systems \cite{breuer}. In the
over thermalization regime, the system is arbitrarily close to the
thermal steady state when it ends the thermal stroke. The term thermal
steady state refers to the state of the machine after it was connected
to the baths (both of them) for a period which greatly exceeds the
thermalization time. Stated differently the thermal steady state is
the fixed point state of the system when it is connected to the baths.
In general it is different from the periodic steady state of the machine
which takes into account also the unitary stroke.

\subsection{Bloch sphere representation of the engine dynamics}

It is highly useful to plot the dynamics of the engine in Fig. 1a
in the Bloch sphere representation. In steady state operation, the
ground state population is fixed in time. The unitary never changes
it, and therefore to maintain steady state periodicity, the thermal
stroke cannot change it either. Thus, the whole dynamics is encapsulated
in the two excited levels and can be most conveniently plotted on
the Bloch sphere as shown in Fig. 1b. The $z$ axis corresponds to
the expectation value of the $\sigma_{z}$ Pauli matrix and since
the interaction is $\sigma_{x}$ (see (\ref{eq: H eng})) the two-level
``spin'' will rotate in the $yz$ plane. The expectation value of
$2\sigma_{z}$ is the population inversion $p_{3}-p_{2}$ and the
expectation values of $\sigma_{y}$ is the y coherence $\frac{1}{2}i(\rho_{23}-\rho_{32}^{*})$
in the energy basis (there is no $x$ coherence in this setup). Therefore
a motion along the $z$ axis is proportional to energy change, and
a motion along the y axis is proportional to coherence change. In
particular the $z$ axis itself indicates zero coherence.

In the over-thermalization regime the thermal stroke must end on the
$z$ axis. Unitary evolution rotates the Bloch vector (green curve)
but cannot change its radius (which is equal to the half the population
inversion). The angular spread of the unitary curve depends on the
interaction strength $\epsilon$, and on the stroke duration. Within
the standard rotating wave approximation a $\pi$ pulse leads to a
$\pi$ rotation, a $\pi/2$ pulse leads to a $\pi/2$ rotation, and
so on. The thermal stroke generates a non unitary evolution (red-blue
arrow). In steady state operation, this brings back the Bloch vector
to the starting point of the unitary stroke and closes the engine
cycle. In the over thermalization regime this starting point is the
thermal steady state.

\subsection{Weakly driven machines $\delta\theta\ll\pi$}

In this paper we shall assume that the units are weakly driven so
that $\theta\to\delta\theta\ll\pi$. Nonetheless, $\delta\theta$
\textit{is finite and not infinitesimal}. Weakly driven machine don't
exhaust their population inversion in each cycle and therefore, in
some sense, don't reach their full potential. Nevertheless, this does
not necessarily means low power since smaller changes in the engine
can be completed in a shorter time.

Clearly by increasing the external driving field, more work can be
extracted. However, \textit{in this work we shall assume that small
$\delta\theta$ engines are given and that $\delta\theta$ cannot
be increased beyond the given value}. The question is what can be
done to increase the work output without adjusting the drive. For
a single unit this is not possible, but we show that for multiple
units that operate collectively it is possible. In practice, $\delta\theta\sim\pi/3$
is small enough to start observing the boost of the collective machine.

Weak drive may result from practical reasons. One reason may be the
lack of strong enough coherent laser source in the needed frequency.
Alternatively the source can be strong but off the atomic resonance.
The large detuning reduced the amplitude of the Rabi oscillation (directly
related to $\delta\theta$). Another motivation to use weak drive
is to avoid excitation of other undesired mechanism and effects such
as nonlinear multiphoton processes.

\section{Coherence extraction and coherence injection}

With the exception of exact permutation, any unitary that operates
on a diagonal state generates coherences in the same basis. This is
especially true for unitaries that are close to the identity where
this is the leading effect generated by the unitary. For diagonal
states and a small time step $dt$, the change in population is proportional
to $dt^{2}$ while the change in coherence is proportional to $dt$.
However, as explained earlier, in the over thermalization regime this
coherence is erased by the baths in the next stroke. 

The idea that facilitates our main results is simple. Instead of letting
the bath erase the coherence we wish to extract and store it elsewhere
for further use. We impose the condition that the extraction does
not change the energy population as this may lead to heat or work
energy exchanges that significantly modify the thermodynamic scheme.
Therefore, the coherence extraction has no energetic role only an
entropic one. A simple way to implement this extraction is to swap
the engine particle with a particle that has the same energy levels
and the same populations but without any coherences. We call this
particle coherence acceptor. This coherence extraction scheme is illustrated
in Fig 2a. Since \textit{the engine operates in the stochastic regime}
(over thermalization), only populations matter, and this swap operation
will have zero impact on the operation of the engine. All energy currents
(heat and work) will remain as they were before the CE. In the weak
coupling limit to the baths (that leads to the Markovian dynamics),
the bath entropy generation is determined by the heat via $\Delta S_{bath}=-Q/T$
($T$ is the temperature of the bath). Therefore $\Delta S_{bath}$
is also indifferent to the presences or lack of coherences at the
beginning of the \textit{thermal stroke}.

\begin{figure}
\includegraphics[width=8.6cm]{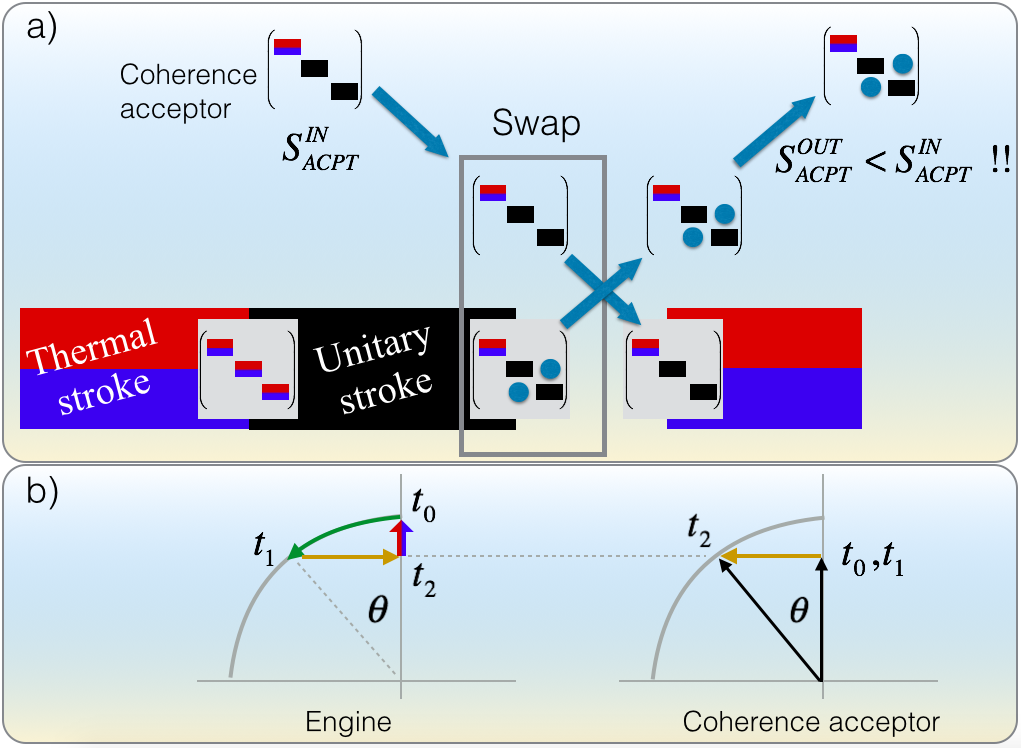}

\caption{Introducing the coherence extraction (CE) process. (a) A density matrix
illustration. The thermal stroke (bottom left of (a)) sets the engine
in a diagonal state. Then, the unitary stroke generates coherences
and some population changes in level 2 \& 3. Next, the engine particle
is swapped with an acceptor particle that has the same energy levels
and the same population. This swap does not involve heat or work exchange.
The engine state becomes diagonal but the engine performance are not
affected by this. On the other hand, the gained coherence of the acceptor
can be used to boost the performance of a different engine that operates
with slightly different temperatures. This shows that even in the
stochastic regime quantum heat engine have quantum byproducts that
can be used. b) Illustration of CE in Bloch sphere representation
(see text).}

\end{figure}

To graphically describe CE two Bloch spheres are needed: one for the
engine and one for the acceptor particle. As shown in Fig. 2b, at
$t_{1}$, just before the thermal stroke, a swap operation starts
between the engine and the acceptor (golden arrows). When it ends
at $t_{2}$, the acceptor has coherence while the engine reaches the
$z$ axis and looses all coherence. Since both particles have the
same populations, the population inversion ($z$ coordinate) is fixed
as well during the swap. As a result, the system energy does not change
and the CE operation does not involve any heat or work exchange between
the system and the auxiliary particle.

Later on coherence extraction will be used together with its complimentary
process, coherence injection, to generate interaction between machines
that lead to collective operation. However, before using CE for constructing
a new type of quantum heat machines, it is worthwhile to discuss some
immediate implications of CE .

\subsection*{Benefits of coherence extraction}

\subsection{Cooling via CE}

Coherence extraction adds coherence to some external particle without
affecting the operation of the engine. What is this coherence good
for? First, in terms of entropy, adding coherences while keeping the
diagonals the same, means that the von Neumann entropy of the particle
has been reduced (the state becomes purer). This is a form of cooling
but without changing the average energy of the particle. Since the
entropy of the auxiliary particle has been reduced and the entropy
production in the baths remained the same, the total entropy production
in the world (see entropy pollution later on) has been reduced. After
we quantify it in Sec. VI it will become clear that this can be a
dramatic effect with important consequences.

\subsection{State activation via CE}

In terms of energy, the added coherence is also important since the
state becomes more active (less passive) than it was before. Passive
states are diagonal states in the energy basis whose populations are
decreasing with energy \cite{alahverdyan04,pusz78,lenard1978Gibbs}.
Stated differently, passive states have no coherences and no population
inversion in the energy basis. All other states are termed 'active
states'. The reason for the names 'active' and 'passive' is the following.
It is \textit{impossible} to extract energy from a passive state by
a unitary transformation if the Hamiltonian is returned to its initial
value after the transformation is completed (e.g. a $\pi$ pulse).
On the other hand, some energy can always be extracted from active
states using unitary operation until they become passive \cite{alahverdyan04}.

Even if the state had some population inversion so that it was not
passive to begin with, adding coherences will make it more active.
To see this, consider a two-step process where in the first step permutations
are used to extract the energy from the population inversion. This
stage extracts work which is equal to the work extracted from the
same original state but with no coherences. However, since permutations
do not alter the magnitude of coherences, the state remains active.
In the second step, a rotation is performed to bring the state to
a diagonal form (whose energy is guaranteed to be lower). Thus the
coherence extraction make the auxiliary state more active and therefore
increases the energy that can be extracted from it. In some cases
the acceptors need to have some population inversion, that is, they
need to be active to begin with. However, one can show that with help
of two $\pi$ pulses it is possible to use a passive acceptor in this
case as well (without an additional energy cost). 

At this point it is clear, at least qualitatively, that CE can cool
and activate a state simultaneously. We point out that in principle,
in other processes, it is possible to cool without activation and
to activate without cooling. Here, however, these two good things
happen simultaneously. Naturally the question that comes to mind is
whether these gifts come for free, or are there some additional consumed
resources? As explained earlier CE in over-thermalization does not
change the heat and work flows of the engine. Nevertheless, there
could be other non energetic resources. The immediate suspects for
extra resources are the acceptor particles, since they have to be
prepared in a specific state. 

In \cite{CollectiveArxiv} we have presented a new way of keeping
track of resources in a very general setup, and showed that the acceptor
particles in CE are not resources but separate objects that perform
a different task from that of the engine. This extra task uses the
coherence byproduct of the engine as fuel. Hence the cooling and activation
are not exactly free, but are carried out without extra additional
resources on top of what the engine is already consuming (heat). The
curios reader is encouraged to follow the logic of resource accounting
in Sec. VIII of \cite{CollectiveArxiv} (v1). However this accounting
is not needed for the collective machine analysis studied here.

Before doing the entropy balance for the CE process let us introduce
another useful process and then do the entropy balance for both processes
together.

\section*{Coherence injection}

The coherence injection (CI) process shown in Fig. 3a is the time-reversal,
complementary process of coherence extraction. A 'coherence donor'
particle with coherence is swapped with the engine particle that has
no coherences. The donor particle and the engine particle have the
same energy levels and the same energy populations. Thus, like CE,
CI does not involve any energy exchange. However, in contrast to coherence
extraction this process is carried out before the unitary stroke.
The injected coherence allows to extract more work for the same unitary
operation. This is easily understood in the Bloch picture. Consider
a unitary rotation with a small angle $\delta\theta\ll\pi$. If there
is population inversion but no coherence, the change in the projection
on the $z$ axis (work) is $1-\cos\delta\theta=\frac{1}{2}(\delta\theta)^{2}+O[(\delta\theta)^{4}]$.
On the other hand, with coherence there will be a linear term as well.
If the maximal amount of coherence is injected \footnote{There are limits because of the positivity constraint of the density
matrix.} before the rotation, then the Bloch vector starts on the equator.
This time the small rotation will lead to an energy change proportional
to $\delta\theta$. Alternatively stated, the injected coherence enables
to get larger change in the populations (z coordinate) for the same
rotation $\delta\theta$.

Another use of coherence injection is to change the character of the
machine without changing the temperatures. As mentioned earlier without
CI the engine in Fig. 1 must satisfy $\Delta E_{c}/\Delta E_{h}\ge T_{c}/T_{h}$
in order to have population inversion and extract work. In Fig. 3b
we show that if coherence is injected, it is possible to extract work
(have an engine) even when there is no initial population inversion.
Once the Bloch vector is not pointing to the south pole, i.e. it is
an active state, it is possible to apply a rotation that will decrease
the energy of the system ($z$ coordinate) and extract work by rotating
the vector towards the south pole. Note that the efficiency is still
given by $1-\Delta E_{c}/\Delta E_{h}$. In this particular engine
CI allows do more work but this is compensated by taking more heat
from the hot bath so the efficiency stays the same. As pointed out
in \cite{scullySingHeat,ScullyLWI}, coherence can be used to extract
energy from a single bath. This is fully consistent with the second
law when taking into account the coherence that has to be supplied
to the machine. 

\begin{figure}
\includegraphics[width=8.6cm]{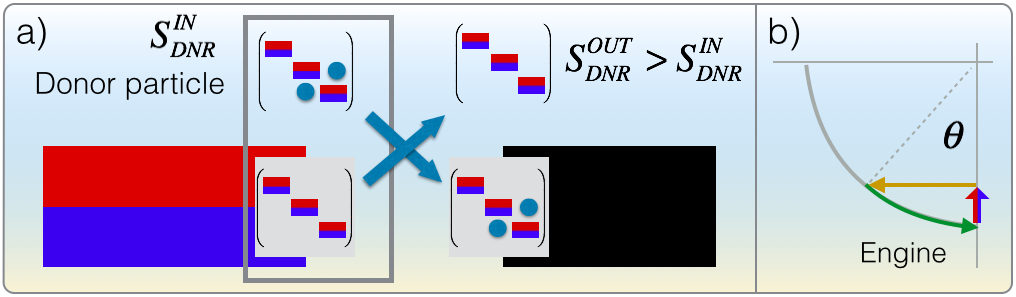}

\caption{(a) Coherence injection (CI). CI is the time inversion of CE. Potentially
it can dramatically boost performance as the unitary stroke can produce
more population change if it starts with some initial coherence. Due
to the complexity of preparing the coherence donor particles, this
process will only be used as a part in a larger device where the donors
are generated automatically by other parts in the device. In fact,
one engine will be used as a donor for a different engine that will
be used as an acceptor. CI can significantly extend the working regime
of engines. (b) Shows an engine operating without population inversion.
This engine can operate also when $T_{c}=T_{h}$ (a single bath) since
the influx of coherence prevents a conflict with the second law. }

\end{figure}

On its own coherence injection is interesting but not very practical.
In CI it is clear that coherence donors are a consumed resource that
has to be accounted for. \textit{Thus, in the present paper CI is
used only as part of a larger device. In this larger device the donors
are created as a byproduct of a different element in the device.}
This scheme is very different from other proposals that use ``coherent
baths'' to supply the machine with coherent (non-thermal) fuel (``phasonium''
\cite{scullySingHeat,ScullyQObook} and ``squeezed baths'' \cite{LutzSqueezedBaths,ParrondoSqueezedBath}).
In \cite{MitchisonHuber2015CoherenceAssitedCooling} it is assumed
that coherence is somehow added to the setup. Coherence injection
is a way to add this coherence without energy costs. 

One may suspect that in such a closed device that has only work and
thermal baths terminals, all benefits of CI and CE will vanish and
the device will operate as a standard thermal machine. However we
show that this is not the case and that the internal CI and CE lead
to observable quantum thermodynamic signatures, and to significant
performance boost.

\section{The second law with acceptors and donors}

Earlier it was qualitatively shown that the engine reduces the entropy
of the acceptor particles (one particle per cycle). In this section
we want to quantify this effect, and show that the scaling properties
of this cooling effect has a very significant consequence to the theory
of heat engines and their reversibility. For this purpose we find
the following two identities extremely useful:
\begin{eqnarray}
S(\rho_{2})-S(\rho_{1}) & = & -tr[(\rho_{2}-\rho_{1})\ln\rho_{1}]-D(\rho_{2}||\rho_{1}),\label{eq: dS 1st id}\\
S(\rho_{2})-S(\rho_{1}) & = & -tr[(\rho_{2}-\rho_{1})\ln\rho_{2}]+D(\rho_{1}||\rho_{2}),\label{eq: dS 2nd id}
\end{eqnarray}
Where $S(\rho)$ is the von Neumann entropy and $D(\rho_{2}||\rho_{1})\triangleq tr[\rho_{2}(\ln\rho_{2}-\ln\rho_{1})]\ge0$
is the quantum relative entropy \cite{VedralRelEnt}. Using (\ref{eq: dS 1st id})
for the acceptors, and (\ref{eq: dS 2nd id}) for the donors we readily
obtain
\begin{eqnarray}
\Delta S_{acpt} & = & -D(\rho_{acpt}^{after}||\rho_{acpt})=-C(\rho_{w}),\label{eq: D coh}\\
\Delta S_{dnr} & = & D(\rho_{dnr}||\rho_{dnr}^{after})=+C(\rho_{dnr}),\label{eq: D dnr coh}
\end{eqnarray}
where $\rho_{w}$ is the density matrix after the unitary stage and
$\rho_{acpt},\rho_{dnr}$ are the state of the acceptor and the donor
before the CE and CI take place. The coherence measure $C$ is given
by $C(\rho)\triangleq D(\rho||diag(\rho))\ge0$. $C$ was identified
in \cite{PlenioCoherence} as a proper coherence monotone. Furthermore,
this form appeared in \cite{LostaglioRudolphCohConstraint}, in the
context of thermodynamic transformations. 

Our first main finding is that the total entropy $\Delta S_{tot}=\Delta S_{baths}+\Delta S_{acpt}+\Delta S_{dnr}$
after one cycle of the machine is
\begin{equation}
\Delta S_{tot}=D(p_{w}||p_{eq}),\label{eq: dS tot D}
\end{equation}
where $p_{w}$ are the populations after unitary stage and $p_{eq}$
stands for the thermal steady state population. The engine entropy
does not appear in the entropy balance because in steady state the
engine state is periodic so after one cycle $\Delta S_{engine}=0$.
Since the relative entropy is always positive (or zero), equation
(\ref{eq: dS tot D}) contains within the 2nd law. However, (\ref{eq: dS tot D})
is stronger than the second law $\Delta S_{tot}\ge0$; the exact equality
can be used to study quantitatively the entropy generation and its
scaling properties. It is interesting that after including the quantum
effects of CE and CI, the second law can be accurately described by
populations only. To prove (\ref{eq: dS aux}) we write:
\begin{eqnarray}
 & \Delta S^{acpt}+\Delta S^{dnr}=\nonumber \\
 & [S(\rho_{w})-S(\rho_{acpt})]+[S(\rho_{eq})-S(\rho_{dnr})]=\nonumber \\
 & S(\rho_{eq})-S(\rho_{acpt})=S(p_{eq})-S(p_{w}),\label{eq: dS aux}
\end{eqnarray}
where we have used the fact that $\rho_{dnr}$ is related to $\rho_{w}$
via unitary transformation and therefore $S(\rho_{dnr})=S(\rho_{w})$.
Using (\ref{eq: dS aux}) and setting $\rho_{1}=diag(p_{w}),\rho_{2}=diag(p_{eq})$
in (\ref{eq: dS 2nd id}):
\begin{eqnarray}
 &  & \Delta S^{acpt}+\Delta S^{dnr}=\nonumber \\
 &  & -\sum_{j}(p_{eq,j}-p_{w,j})\ln p_{eq,j}+D(p_{w}||p_{eq})=\nonumber \\
 &  & -\sum_{j}(p_{eq,j}-p_{w,j})(-\Delta E_{j}/T_{j}+Z)+D(p_{w}||p_{eq})=\nonumber \\
 &  & Q_{h}/T_{h}+Q_{c}/T_{c}+D(p_{w}||p_{eq}),\label{eq: ds dnr acp}
\end{eqnarray}
where $Z$ is a normalization factor that comes from $p_{eq}$. Since
the first two terms on the right hand side of (\ref{eq: ds dnr acp})
are equal to $-\Delta S_{baths}$, (\ref{eq: ds dnr acp}) immediately
leads to (\ref{eq: dS tot D}). 

To appreciate the dramatic implication of (\ref{eq: dS tot D}), we
study entropy generation in engines where the unitary operation only
slightly changes the populations (very small rotation) with respect
to the thermal steady state. Let us denote the population change as
$\delta p=p_{w}-p_{eq}$. the expression for the baths entropy is:
\begin{equation}
\Delta S_{baths}=-\frac{Q_{h}}{T_{h}}-\frac{Q_{c}}{T_{c}}=-\sum\delta p_{j}\ln p_{eq,j}.\label{eq: S bath dp}
\end{equation}
For small $\delta p$ the expansion of the relative entropy is:
\begin{equation}
\Delta S_{tot}=D(p_{w}||p_{eq})=\sum_{j=1}^{3}\frac{1}{2}\frac{\delta p_{j}^{2}}{p_{eq,j}}-\frac{1}{6}\frac{\delta p_{j}^{3}}{p_{eq,j}^{2}}+O(\delta p^{4}).\label{eq: D expand}
\end{equation}

$\Delta S_{baths}$ contains a linear order in $\delta p$ but as
we see from (\ref{eq: D expand}), $\Delta S_{tot}$ is only quadratic
in $\delta p$. This means that in coherence extraction, a remarkable
cancellation of the entropy generation leading order $O(\delta p)$
(\ref{eq: S bath dp}) takes place. Thus, the cooling effect of the
acceptors is not a minor effect even with respect other entropy modifying
elements (baths). It is a leading order effect and as we shall see
in the next sections, it can lead to new types of machines and to
new reversibility limits. Another important property of (\ref{eq: dS tot D})
is that it holds regardless if there is a donor or not (acceptor is
necessary). However, the magnitude of $\delta p$ is affected by the
donors coherence as explained earlier.

\section{Entropy pollution and a new reversible limit for Otto engines}

Entropy generation gives some information about reversibility. However,
to a certain extent, only the limit of zero entropy generation is
perfectly clear. In particular, entropy generation alone is not sufficient
for comparing the reversibility of two engines. For example, one engine
has $\Delta S_{tot}=1$ and work $W=1$ and the other has $\Delta S_{tot}=1$
and $W=1000$. Clearly, the second engine is more reversible. To produce
the $W=1000$ as the second engine a thousand engines (or thousand
more cycles) of the first kind are needed. The total entropy production
of a thousand engines of the first kind is $\Delta S_{tot}=1000$,
which is clearly less reversible compared to the second engine with
$\Delta S_{tot}=1$ for the same amount of work. This motivates us
to define the entropy pollution ($\text{EP}$) irreversibility measure
as:
\begin{equation}
\text{EP}\triangleq\frac{\Delta S_{tot}}{W_{tot}},\label{eq: EP def}
\end{equation}
where the quantities are calculated for a whole cycle in steady state.
Note that in \cite{Woods2015EffEng} the same measure was used to
quantify the charging quality of a battery. Here, however, we use
(\ref{eq: EP def}) to quantify the reversibility of the whole setup
including all the elements. One might be tempted to suggest $\eta_{c}-\eta$
as a measure of irreversibility, but the problem is that when there
are multiple baths as in the collective machine the Carnot efficiency
is not uniquely defined. Depending on the bath connectivity there
could be many different Carnot machines with different efficiencies.
In contrast, the definition (\ref{eq: EP def}) is free from this
ambiguity.

For classical or quantum engines \textit{without }coherence injection
or extraction:
\begin{eqnarray}
\text{EP\ensuremath{{}_{\cancel{CI\:CE}}}} & = & -\frac{-(W-Q_{h})/T_{c}+Q_{h}/T_{h}}{W}\nonumber \\
 & = & \frac{(1-\eta)/T_{c}-1/T_{h}}{\eta}=\frac{1}{T_{c}}\frac{\eta_{c}-\eta}{\eta}.\label{eq: EP no CE}
\end{eqnarray}
For the Carnot efficiency the entropy pollution (\ref{eq: EP no CE})
yields zero as expected. This general expression for $\text{EP\ensuremath{{}_{\cancel{CI\:CE}}}}$
shows that as long as the efficiency and temperatures are kept the
same, the entropy pollution is the same. It does not matter if there
are multiple units operating independently in parallel or if the value
of $\delta p$ is changed. Just as an example, for the Otto engine
in Fig. 1 $\eta=1-\frac{\Delta E_{c}}{\Delta E_{h}}$ so we verify
that without CE the entropy pollution does not depend on $\delta p$
(i.e. $O(\delta p^{0})$). Next we study the case of an engine with
CE. Using (\ref{eq: D expand}) we find

\begin{equation}
\text{EP\ensuremath{{}_{CI\:CE}}}=\frac{D(p_{w}||p_{eq})}{\sum_{j=1}^{3}E_{j}\delta p_{j}}=O(\delta p).\label{eq: EP with CE}
\end{equation}
where we used $W=\sum_{j=1}^{3}E_{j}\delta p_{j}=O(\delta p)$ and
(\ref{eq: D expand}). This result has a profound implication as we
illustrate next. Let our goal be the production of $W_{0}$ work using
one engine. This $W_{0}$ is associate with some population change
per cycle $\delta p_{0}$. Now instead of doing $\delta p_{0}$ per
cycle we only do $\delta p_{0}/N$ per cycle and perform $N$ time
more cycles compared to the previous scenario. Clearly the same amount
of work $W_{0}$ is obtained in both scenarios. Yet, in the smaller
steps case, the engine operates closer to the thermal steady state
$p_{eq}$. Although this is similar in spirit to a quasi-static process,
it is easy to verify from (\ref{eq: EP no CE}) that \textit{without
CE, the entropy pollution is not affected by this splitting to small
steps}. Even though the system is now always closer to the thermal
steady state, the irreversibility (entropy pollution) has not decreased.
In stark contrast, according to (\ref{eq: EP with CE}) with CE we
get 
\begin{equation}
\text{EP\ensuremath{{}_{CE}}}=\frac{N\:O(\delta p_{0}^{2}/N^{2})}{W_{0}}=O(1/N).
\end{equation}
By taking the quasi-static limit $N\to\infty$ we now obtain that
$\text{EP\ensuremath{{}_{CE}}}\to0$. This is a reversible limit as
no entropy generated in the total setup (acceptors+baths+engine).
At first, this result seems strange. This Otto engine produces finite
work%
{} and its efficiency is less than Carnot ($\eta=1-E_{c}/E_{h}$), and
yet the entropy pollution is zero, i.e. it is reversible. This seems
odd as the Carnot's theorem states that two-bath reversible machines
must operate at Carnot efficiency. Indeed, without CE the second law
permits only one type of reversible machine - the Carnot machine.
However, \textit{the inclusion of CE changes the entropy balance and
enables the existence of other types of reversible machines whose
efficiencies are lower than that of Carnot}. In fact, when expanding
the acceptors entropy for small $\delta p$ one finds that the bound
predicted by the second law $\Delta S_{tot}=\Delta S_{bath}+\Delta S_{acpt}=0$
leads in the limit $\delta p\to0$ to the observed efficiency $1-\Delta E_{c}/\Delta E_{h}$
and not to the Carnot efficiency. Once again we have full consistency
with the second law.

\begin{figure}
\includegraphics[width=8.6cm]{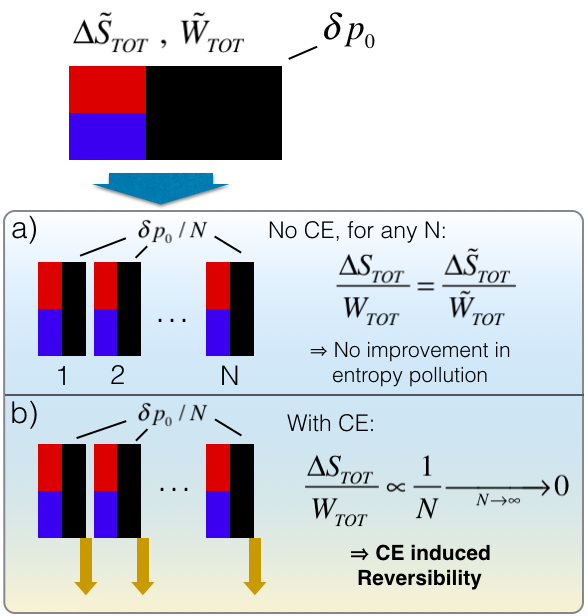}

\caption{(Top) Splitting the work done in one cycle $\tilde{W}_{tot}$ into
$N$ many cycles, makes the system stay closer to the thermal steady
state. However, as shown in (\ref{eq: EP no CE}) the entropy pollution
irreversibility measure is not reduced or affected by this splitting
(see (a)). In contrast, when CE is added to the scheme (see (b)),
the entropy pollution goes to zero in the limit of large $N$. This
enables to achieve reversibility even though the efficiency is lower
than the Carnot efficiency. As explained in the text this is fully
consistent with the second law. }

\end{figure}

Note that this reversible limit does not have to be slow. Instead
of doing $N$ times more cycles, it is possible to have $N$ engines
working in parallel for the original number of cycles. Other aspects
of reversibility will be discussed in a different publication.

In the $\delta p\to0$ limit entropy is not generated. It is simply
redistributed between the baths and the acceptors. In the scenario
studied above the heat and work flows were not affected by the coherence
extraction. To avoid unnecessary entropy generation coherence was
stored in the acceptors that act as coherence repositories. Although
this is interesting from an informational point of view, our main
goal in this paper is to show new ways of harnessing coherence to
boost performance. Our approach to achieve this performance boost
is to construct a collective operation scheme as discussed in the
next section.

\section{Collective coherent engines}

As we have shown, CE in quantum heat engines reduces the entropy production
but it does not affect the power performance of the engine. It enables
other tasks like cooling and activation for free, but the main function
of the engine, power production, does not benefit from the free coherence
production. Now we present a scheme where CE and CI are used to simultaneously
boost power (not efficiency) and reversibility. This scheme has another
very appealing property; \textit{it does not use any external acceptor
and/or donor particles as before}. It is based solely on coherence
sharing between engines. When two engines interact one engine perform
as donor and the other as an acceptor. 

Earlier we introduced the coherence injection process. Despite the
potential power boost it offers, the preparation of a stream of coherent
donor particles makes the scheme less appealing. Fortunately, as shown
earlier, engines produce coherence that can be harvested and used
without any energetic cost. It is natural, then, to use one engine
as a donor and pass its coherence to a second engine that will serve
as an acceptor. The first engine experiences coherence extraction
and the second engine experiences coherence injection. The second
engine gets a power boost by starting the unitary with some coherences
as explained earlier and in \cite{EquivPRX}. The total power will
be the power obtained from the two engines (the first stage also produce
power). This process can be repeated for multiple stages until the
is no more coherence to harvest from the last stage. In what follows
we study two different regimes and compare to individual operation
without coherence sharing.

\subsection{Quadratic power boost for collective machines }

Our collective coherent engine scheme is shown in Fig. 4a. The first
engine donates its coherence to the second stage engine. The temperatures
of the second engine baths are chosen so that resulting thermal steady
constitutes an acceptor for the first engine. It is a perfect synergy
where one stage prepares what the other stage needs while all units
produce work at the same time. In addition, later on in the cycle
the second stage engine donates its coherence to a third-stage engine,
and so on. Each donor-acceptor interaction between the engines is
carried out by a full swap of the two engine particles. Thus only
engine particles are participating in this scheme and there are no
external donor-acceptor particles. The question of CI extra resources
is irrelevant in this device.

Due to this process of coherence transfer, each engine continues the
unitary evolution where the previous stage has stopped (green line
in Fig 4a and 4b). Thus, instead of many copies of $\delta\theta$
rotations as happens without CE-CI, the whole system performs $\delta\theta$
unitary as shown in Fig 4b. \textit{It is like an assembly line where
each engine does its part}. %
{} We denote by $\Omega=Nd\theta$ the collective angle. For initial
population inversion $dp_{0}$ the total change in the population
inversion is $\delta p=dp_{0}(1-\cos\Omega)=2dp_{0}\sin^{2}\frac{\Omega}{2}$
and therefore the work of the collective device is
\begin{equation}
W_{coll.}=(E_{h}-E_{c})dp_{0}\sin^{2}\frac{\Omega}{2}.\label{eq: Wcoll}
\end{equation}
Next, we wish to compare the collective operation to the operation
of $N$ independent units with the same $\delta\theta$. Now that
there is no CE and CI there is no need for using different temperatures
for different engines. Nevertheless, for the fairness of comparison,
we do not allow the standalone units to be connected to any cold and
hot temperatures. The temperatures must be chosen from those available
to the collective machine. This principle will be used in other comparisons
later on. 

We choose the unit that produces the maximal amount of work when working
in a standalone mode, and set the other $N-1$ units to work with
the same temperatures (N independent copies of the same engine-bath
setup). The engine that produces the maximal amount of work as a standalone
unit is the first one, because it has the highest initial population
inversion. Thus the total power in the standalone work optimized (SWO)
case is
\begin{equation}
W_{N,swo}=N(E_{h}-E_{c})dp_{0}\sin^{2}\frac{\Omega/N}{2}.\label{eq: W_Nswo}
\end{equation}
The work boost for $\Omega\le\pi$:
\begin{equation}
\frac{W_{coll.}}{W_{N,swo}}=\frac{\sin^{2}\frac{\Omega}{2}}{N\sin^{2}\frac{\Omega/N}{2}}\ge\frac{\sin^{2}\frac{\Omega}{2}}{(\frac{\Omega}{2})^{2}}N\ge\frac{4}{\pi^{2}}N.\label{eq: work boost}
\end{equation}
When $\Omega\ll\pi$ the collective work boost is simply $N$ (the
coefficient is unity and not $4/\pi^{2}$). Alternatively stated,
\textit{while in the standalone case the work for a fixed $d\theta$
scales like the number of units $N$, in the collective case the work
scales like $N^{2}$}.

We point out that in the performance comparison studied in this paper
the 'unitary strength', $\delta\theta$, of each units is fixed. We
do not compare machines with different values of $\delta\theta$.
Instead, we ask the following question: given that $N$ machine with
fixed $\delta\theta$ are available, is it possible to boost their
performance by making them work collectively? 
\begin{figure}
\includegraphics[width=8.6cm]{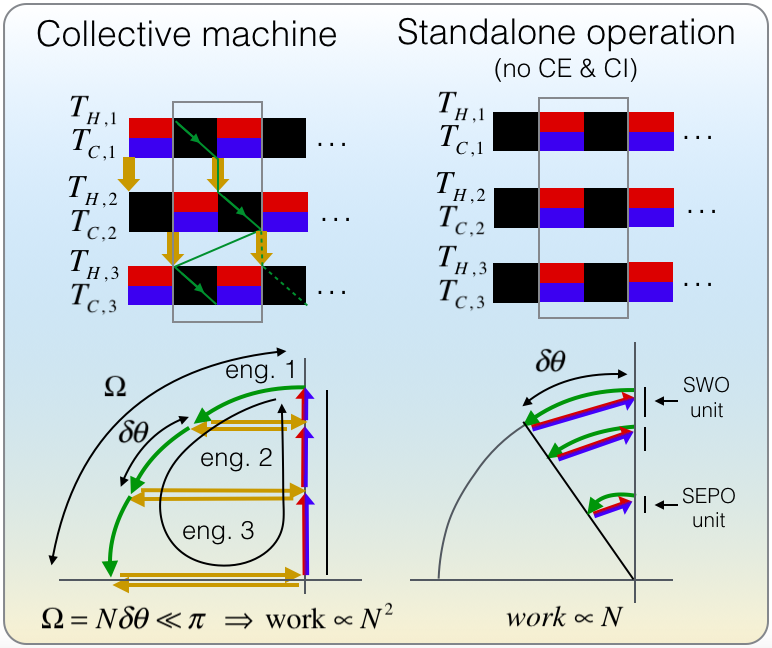}

\caption{(left) In our collective heat machine scheme coherence is extracted
from one machine and injected to the other. The resulting ``collective
unitary'' of angle $\Omega$ produces more than $N$ times more work
compared to the standalone operation (right). The work boost is $N$
even when comparing to N standalone units that produce the maximal
amount of work (SWO units). This means that for a fixed $\delta\theta$
\textit{the collective work scales quadratically with the number of
units rather than linearly as in the standalone operation. }}

\end{figure}

\subsection{Saturation and boosted linear scaling of collective machines}

The expression (\ref{eq: Wcoll}) for the collective work is bounded
and reach its peak value at $\Omega=\pi$. On the other hand, the
expression for the non interaction SWO units always scales like N.
Does this means that for some large enough value of $N$ the collective
machine will produce less power than $N$ SWO machine? not quite.
Despite the saturation, collective operation still give rise to significant
power boost.

Let $M=\pi/\delta\theta$ be the number of units needed for a $\Omega=\pi$
collective rotation. We have at our disposal $N\gg M$ units, each
of strength $\delta\theta$ (which is fixed and $N$ independent).
There is no point in letting more than $M$ units interact together.
For example $2M$ units operating collectively will generate a $2\pi$
rotation that amounts to zero work. Hence, we divide the units into
sets of $M$ units (one set may contain less than $M$ units). Each
set is operated as a collective machine. For the complete sets it
holds that $\Omega=\pi$, so according to (\ref{eq: Wcoll}) their
power amounts to $(E_{h}-E_{c})dp_{0}\left\lfloor N/M\right\rfloor $
where $\left\lfloor \cdot\right\rfloor :=Floor(\cdot)$. Adding the
power from the single incomplete set we get
\begin{eqnarray}
 &  & W_{coll,N>M}=\nonumber \\
 &  & (E_{h}-E_{c})dp_{0}\{\left\lfloor N/M\right\rfloor +\sin^{2}(Rem(N,M)\delta\theta/2)\}\nonumber \\
\label{eq: WcolNM}
\end{eqnarray}
where $Rem$ is the remainder. For large $N$ and small $\delta\theta$
($\delta\theta\le\sim\pi/5$ or $M\ge\sim5$):
\begin{equation}
\frac{W_{coll,N\gg M}}{W_{1,SWO}}\simeq\frac{N/M}{\delta\theta^{2}/4}=\frac{4M}{\pi^{2}}N\label{eq: LinBoost}
\end{equation}
Comparing this result to the SWO scaling $\frac{W_{N,SWO}}{W_{1,SWO}}=N$
we observe that there is a $\frac{4M}{\pi^{2}}$ power boost even
though the behavior is no longer quadratic for $N\ge M$. This is
illustrated in Fig. 6 for $M=20$. The graph aim to show the $N\ge M$
behavior. The red dot show the actual power of the collective machine
as give by expression (\ref{eq: WcolNM}). Initially the power grows
quadratically (blue curve). After that the power follows the boosted
linear scaling (green curve) given by (\ref{eq: LinBoost}). For comparison,
the orange curve shows the linear power growth of non interacting
units. For smaller $\delta\theta$ the quadratic scaling regime is
larger. 

\begin{figure}
\includegraphics[width=8.6cm]{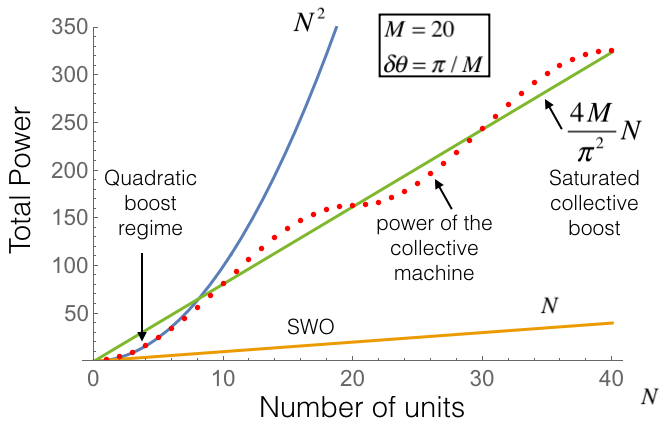}

\caption{When the number of units $N$ exceeds the number of units needed to
complete a $\pi$ rotation in the Bloch sphere $M$, the boost changes
from quadratic (blue) to linear (green). Nevertheless the linear scaling
of the collective machine is still boosted by a factor $4M/\pi^{2}$
(for $M\ge\sim5$) compared to the standard linear scaling $N$ (orange)
of the non interacting SWO units. Depending on $\delta\theta$, the
quadratic regime can be larger.}
\end{figure}

\subsection{Extraordinary reversibility via symbiotic relations at $\Omega=\pi$
collective angle }

In this section we study the intriguing case of $N\delta\theta=\pi$.
In terms of work boost we simply need to set $\Omega=\pi$ in (\ref{eq: work boost}).
However, in this section we are interested in the entropy pollution
of the collective device. Surprisingly, we find that the collective
machine can be more reversible than the most reversible unit in the
collective. Moreover, in this machine work yield boost and reversibility
boost are not in conflict. Both of them increase with the number of
units. This is due to a unique symbiotic mechanism that takes place
in the $\Omega=\pi$ collective machine.

In the $\Omega\ll\pi$ case all units are close to the north pole,
so the entropy pollution of the collective machine is very similar
to that of a standalone machine with temperature that corresponds
to maximal population inversion (north pole of the Bloch vector). 

In addition, in the $\Omega<\pi$ collective machines there is some
residual unused coherence since the Bloch vector does not return to
the $z$ axis at the end of the collective unitary ($\Omega$ rotation).
Since there is no external acceptor to store this coherence it is
being erased by the baths. Consequently, the bath entropy generation
is larger compared to the case where coherence is stored.

For a fair performance evaluation of the $\Omega=\pi$ engine, it
is important to take the right comparison reference. We shall make
two different comparisons. First we compare the $\Omega=\pi$ case
to the standalone work optimized case studied in the previous section.
Next we shall remove the work optimization constraint and compare
to the most reversible unit in the collective. First we start with
the scaling of the entropy pollution in the $\Omega=\pi$ case. 

\begin{figure}
\includegraphics[width=8.6cm]{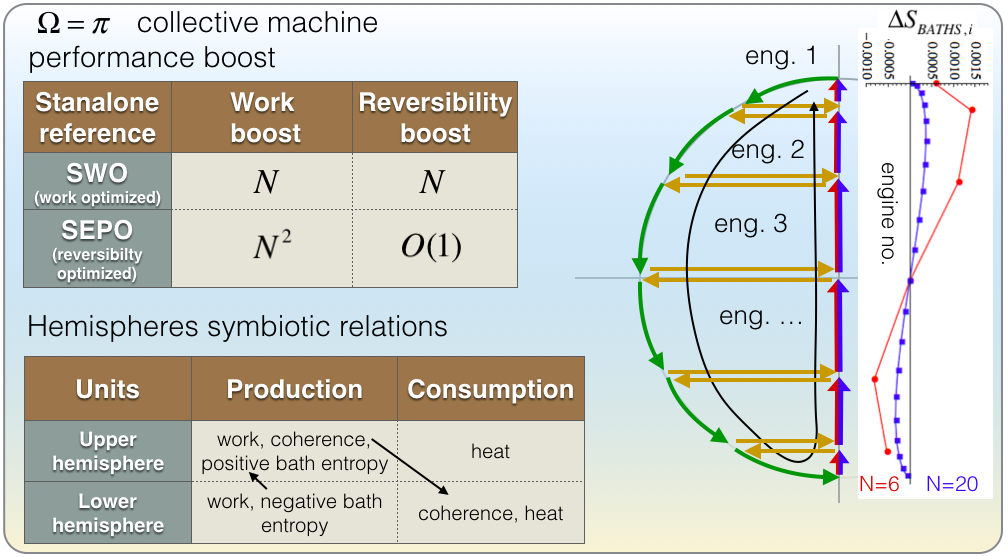}

\caption{The $\Omega=\pi$ collective machine shows a remarkable symbiotic
interaction between units in the upper and lower hemisphere. The coherence
the lower units require in order to operate as engines (they have
no population inversion) is supplied by the upper units. In return
the lower units produce negative bath entropy that in the limit of
large $N$ compensates for the positive entropy generation of the
upper units. The entropy generation in the baths of each units is
shown in the rightmost graph. In the limit of large $N$ the graph
becomes antisymmetric and the total entropy production goes to zero.
In this limit the $\Omega=\pi$ is reversible. }
\end{figure}

\subsection{Entropy pollution of the $\Omega=\pi$ machine}

In our collective scheme all units but the last one, have acceptors
(the next stage engine) that remove the remaining coherence after
each engine complete its unitary stroke. Hence, in the general case,
the entropy generation in all units but the last can be evaluated
using (\ref{eq: dS tot D}). However, when $\Omega=\pi$ the last
machine ends at the south pole with no coherences so there is nothing
to extract. Thus, in the $\Omega=\pi$ case we can safely use (\ref{eq: dS tot D})
for all engines including the last one, and get that the total entropy
production is
\begin{eqnarray}
\Delta S_{tot} & = & \Delta S_{baths}^{\Omega=\pi}=\sum_{i=1}^{N}D(p_{eq}^{(i+1)}||p_{eq}^{(i)}),\\
p_{eq}^{(i)} & = & \{1,e^{-\frac{\Delta E_{c}}{T_{c,i}}},e^{-\frac{\Delta E_{h}}{T_{h,i}}}\}/(1+e^{-\frac{\Delta E_{c}}{T_{c,i}}}+e^{-\frac{\Delta E_{h}}{T_{h,i}}}),\nonumber \\
\end{eqnarray}
where we used $p_{eq}^{(i+1)}=p_{w}^{(i)}$. This $p_{eq}^{(i+1)}=p_{w}^{(i)}$
condition determines the temperatures of the $i+1$ baths - it is
the basic population matching in CE and CI that assures that there
are no energy exchanges between the units. In addition, $\Delta S_{tot}=\Delta S_{baths}$
follows from the fact that in this system there are no external particles
so in a periodic process the total entropy of the elements can only
be modified in the baths. For, small population change $\delta p$
associated with small $\delta\theta$, we can write: $D(p_{eq}^{(i+1)}||p_{eq}^{(i)})=\frac{1}{2}\sum_{j=2}^{3}\frac{\delta p_{i}^{2}}{p_{eq}^{(i)}(j)}$
and $\delta p_{i}=dp_{0}\sin\theta_{i}\delta\theta$, where $j$ is
a level index and $i$ is an engine index. Using $d\theta=\pi/N$
we have $\Delta S_{baths}=\Delta S_{tot}=\sum_{i=1}^{N}\frac{1}{2}\sum_{j=2}^{3}\frac{\delta p_{i}^{2}}{p_{eq}^{(i)}(j)}=\sum_{i=1}^{N}\frac{1}{2}\sum_{j=2}^{3}\frac{dp_{0}^{2}\sin^{2}\theta_{i}\delta\theta\frac{\pi}{N}}{p_{eq}^{(i)}(j)}.$
Since $p_{eq}^{(i)}(j)$ is a function of $\theta_{i}$ we get $\Delta S_{baths}=\Delta S_{tot}=\frac{1}{N}\sum_{i}F(\theta_{i})\delta\theta$.
The sum can be approximated by an integral for large N. Since the
integral does not depend on N we get that $\Delta S_{tot}^{\Omega=\pi}\propto1/N$.
The total work in this case is $dp_{0}(E_{h}-E_{c})$ regardless of
$N$, so finally, the entropy pollution scaling is: 
\begin{equation}
EP{}_{\Omega=\pi}\propto1/N.\label{eq: EP_pi}
\end{equation}

\subsubsection*{Comparison to the standalone work optimized case}

We saw that if the N units that constitute the collective engine,
work independently and the temperatures are chosen to maximize the
work, their total work is $N$ times weaker compared to the collective
machine (see (\ref{eq: work boost})). One might suspect that this
enhancement is accompanied by degradation of reversibility, i.e. in
an increased entropy pollution. In the standalone work optimized case
the entropy pollution is the same for all units. It is given by (\ref{eq: EP no CE})
with the temperature and Carnot efficiency that fits the north pole
of the Bloch sphere. The exact expression is not important. What matters
is that $EP_{swo}$ does not depend on the number of units $N$. Using
(\ref{eq: EP_pi}) we immediately get 
\[
\frac{EP{}_{\Omega=\pi}}{EP_{swo}}=\frac{O(1/N)}{O(N^{0})}=O(1/N)
\]
\textit{The conclusion is that the $\Omega=\pi$ collective machine
produce $N$ times more work, and at the same time it is $N$ times
more reversible compared to the standalone work optimized case}. In
addition this shows that \textit{even though there are no external
auxiliary particles, CE and CI are still highly useful as they can
increase work yield and reversibility}.

Note that the comparison above is very strict. If the coherence exchange
is not performed but the temperature of each unit is the same as in
the collective machine, then all the lower hemisphere units will turn
into refrigerators and consume work instead of producing work. As
a results, the total work output in this 'no CI CE' case becomes zero
for large $N$ (the cancellation is not perfect for small values of
$N$). Of course in such a case there is little motivation to use
units that do not produce work. This is why we chose to compare it
to the power-maximized standalone case. Next, we compare it to the
standalone $\text{EP}$ optimized case.

\subsubsection*{Comparison to the most reversible unit}

If we look on the individual units operating at different temperatures
needed for the collective machine, the unit with the lowest entropy
pollution is unit number $\left\lceil N/2\right\rceil :=\text{Ceiling}(N/2)$,
the one that starts just above the equator \footnote{As explained earlier when the machines do not interact but still work
with the baths temperature used for the collective machine, then the
units in the lower hemisphere no longer operate as an engine.}. This unit has the lowest entropy pollution since for large $N$
its Carnot efficiency approaches the actual efficiency $\eta_{c}(\left\lceil N/2\right\rceil )\to\eta=1-\Delta E_{c}/\Delta E_{h}$
(see (\ref{eq: EP no CE})). Now we wish to compare the collective
machine to the most reversible unit (in standalone operation) in the
collective. Following the same logic as in the SWO case, $N$ copies
of this most reversible unit are chosen as a comparison reference.
We denote this reference by SEPO for 'standalone EP optimized'. We
once again point out that for the fairness of comparison, we are only
allowed to use bath temperatures that are available to the collective
machine. It is simple to show that this time the ratio between the
work of the collective $\Omega=\pi$ machine and the SEPO scales like
$N^{2}$. However, when comparing the entropy pollution of the two
scenarios one finds something surprising.

After some algebra one can show that the entropy pollution of the
SEPO scales like $1/N$, just like the $\Omega=\pi$ collective entropy
pollution (\ref{eq: EP_pi}). In Fig. 8 the ratio $\text{EP}_{coll.}/\text{EP}_{sepo}$
is plotted for different minimal temperatures. As expected from the
scaling arguments above, the ratio saturates for large $N$. The most
striking feature in this figure is that the collective machine can
be more reversible (less polluting) than the most reversible unit
(the plotted ratio is smaller than one). The second feature that stands
out in Fig. 8 is the dependence on the parity of units number. This
parity dependence is not coming from $\text{EP}_{coll.}$ but from
the SEPO reference. For even $N$ the most reversible machine is at
an angle $\frac{\pi}{N}$ from the equator while for odd $N$ the
angle is $\frac{\pi}{2N}$. For large $N$ this means a factor of
two in the $\eta_{c}-\eta$ factor that appears in (\ref{eq: EP no CE}).
 The figure shows that when the cold temperature is comparable (or
higher as we see numerically) to the cold gap, the collective machine
is more reversible than the SEPO regardless of parity. When the low
temperature is below the cold gap we observe that only the even $N$
values lead to superior reversibility of the collective machine.

Of course this small improvement shown in Fig. 8 is not a performance
boost. The performance boost is in the power as explained in the beginning
of this section. Nevertheless, it is remarkable that the collective
machine can be more reversible than the most reversible unit - even
by a little. 

The two comparisons shown in the upper table in Fig. 7, can be combined
by considering the figure of merit $W/\mbox{EP}$ that can be understood
as work multiplied by reversibility measures $1/\text{EP}$. Repeating
the same argument as before it is easy to verify that the $W/\mbox{EP}$
measure of the $\Omega=\pi$ machine is $N^{2}$ better compared to
both SWO and SEPO standalone operation. 

\begin{figure}
\includegraphics[width=8.6cm]{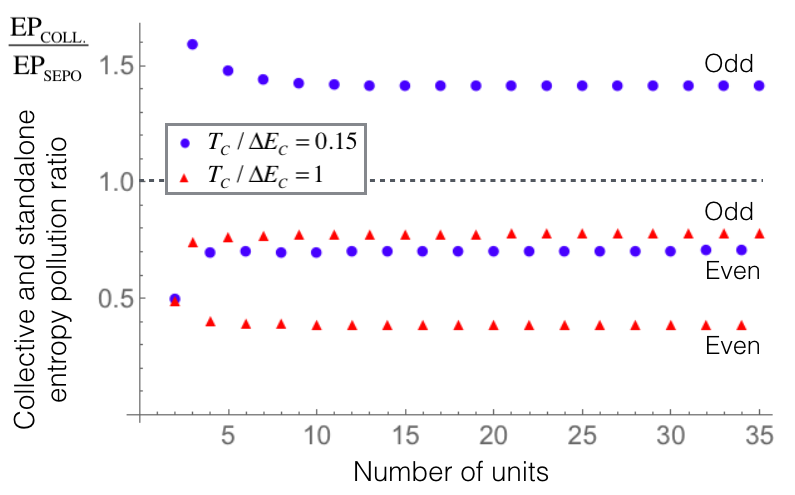}

\caption{Comparing the entropy pollution of the collective machine to its most
reversible element in standalone units (SEPO). \textit{The graph shows
that the reversible collective machine can be more reversible than
the most reversible unit that composes it} (pollution ratio smaller
than one). The parity dependence is explained in the text. The collective
is not always more reversible as shown by the blue circles that correspond
to a collective machine with colder baths.}

\end{figure}

In what follows we present a different point of view that unravels
the interesting dynamics that leads to high reversibility and high
work yield in the $\Omega=\pi$ collective machine. The rightmost
graph in Fig. 7 shows the bath entropy generation $\Delta S_{baths,i}=-Q_{h}^{(i)}/T_{h}^{(i)}-Q_{c}^{(i)}/T_{c}^{(i)}$
of each engine (red circles $N=6$, blue squares $N=20$). In the
lower hemisphere there is a negative entropy production. Clearly,
from (\ref{eq: ds dnr acp}) this becomes possible only if there is
CI. The source of the injected coherence are the engines in the upper
hemisphere. As the number of engines increases the graph becomes more
antisymmetric and the total sum of entropy generation in the various
baths becomes closer to zero. The reason for the almost exact, leading
order, cancellation of the $i$ baths entropy generation with that
of $N+1-i$ baths (antisymmetry), can be understood in the following
way. Equation (\ref{eq: D coh})-(\ref{eq: dS tot D}) lead to
\[
\Delta S_{baths,i}=C(\rho_{dnr,i})-C(\rho_{w,i})+D(p_{w,i}||p_{eq,i}).
\]
The last term scales like $1/N^{2}$, so the leading term is the coherence
difference that scales like $1/N$. Since up to level permutation
$\rho_{dnr,i}$ is equal to $\rho_{w,N+1-i}$ it follows that $C(\rho_{dnr,i})=C(\rho_{w,N+1-i})$
and therefore
\begin{eqnarray}
\Delta S_{baths,i}+\Delta S_{baths,N+1-i} & = & D(p_{w,i}||p_{eq,i})\nonumber \\
 & + & D(p_{w,N+1-i}||p_{eq,N+1-i})\nonumber \\
 & = & O(1/N^{2}).\label{eq: dS pairs}
\end{eqnarray}
This pairwise cancellation of the leading order, $1/N$, in the bath
entropy generation, enables the reversible limit of the $\Omega=\pi$
machine. After providing the mathematical explanation for this pairwise
cancellation, it instructive to examine the unique physical circumstances
that enables this result. The pairwise cancellation is a manifestation
of a very interesting symbiotic relation between the engines in the
upper hemisphere and the engines in the lower hemisphere. The upper
units and lower units operate in a very different manner. The upper
ones produce work and coherence. The engines in the lower hemisphere
also produce work but since they do not have population inversion,
they must use coherence in order to produce work (see Fig. 3b). The
coherence they need comes from the engines in the upper hemisphere.
This shows what the lower hemisphere units gain from the upper ones.
What about the other direction? Do the upper ones get something in
return from the lower ones? The answer is yes. As explained earlier,
and as shown in the graph in Fig. 7), due to the coherence injection
the engines in the lower hemisphere have\textit{ negative} bath entropy
generation. Equation (\ref{eq: dS pairs}) implies that for large
$N$ this negative contribution compensates the positive bath entropy
generation of the upper hemisphere engine. In summary, the upper units
'sponsor' the lower units, while the lower units clean up after the
upper ones (see lower table in Fig. 7). To the best of our knowledge
this is the first example of symbiotic relations between different
'species' of machines in thermodynamics.

If $N$ is large enough this machine operates as a refrigerator when
the order of the swap operations is reversed. With some modifications
that will be described elsewhere, the reverse operation can be improved. 

Note that there is no Carnot theorem for multiple baths. There are
efficiency limits but for the same multiple baths there are different
reversible machines that operate at different efficiencies. Only in
the two bath case the reversibility constraint is sufficient to fix
the efficiency. Hence, \textit{the $\Omega=\pi$ heat engine has a
reversible limit with efficiency $\eta=1-\Delta E_{c}/\Delta E_{h}$
that is not fixed by the baths temperatures. }

\section{Concluding remarks}

In this work we have introduced two quantum thermodynamics operations
- coherence extraction (CE) and coherence injection (CI). The first
(CE) enables to reduce the entropy pollution while the latter can
boost performance and extend the parameter regime in which the machine
operates as an engine. By allowing heat machines to interact with
each other according to these two interactions, we studied what is,
to the best our knowledge, the first example of coherent collective
heat machine. For certain parameters the work in this device scales
quadratically with the number of composing engines and on top of the
work boost it can also be more reversible compared to standalone operation.
In addition, the collective machines constructed here make use of
coherence extraction and injection without external coherence storage.
The terminals of the collective machine are just thermal baths and
a classical driving field (work repository). Nevertheless the internal
coherence manipulation enable to extract more energy from the heat
baths compared to individual standalone operation.

The harvesting of coherence has a significant impact on the role of
the second law in heat machine operation, as less information is erased
in each cycle. This is neatly expressed in a simple relative entropy
expression for the total entropy production. The relative entropy
scaling shows that the information salvaged in the coherence can be
significant enough to enable new reversible limits for otherwise irreversible
machines. 

For a standard quantum machine, strong thermalization destroys coherence
and turns the device into a stochastic thermal machine. The present
study breaks this seemingly necessary connection between full thermalization
and stochastic-incoherent operation. In the collective operation mode,
full thermalization and coherence effects coexist. 

The use of full swap for CE and CI was chosen to avoid entanglement
generation between particles as it may increase the entropy of the
reduced systems and increase the entropy pollution. Nonetheless, it
is possible that entanglement created by partial swap can be put to
use in some variation of the engines studied here.

CE and CI give rise to new types of quantum thermodynamic devices.
Experimental realizations of the ideas presented here call for techniques
developed for quantum computation in order to implement the swap gates.
In addition, open system techniques are needed to generate the thermalization
stroke. Although these techniques are not yet perfectly integrated
together at the moment, both are feasible with present day technology.
In particular, it seems likely that the rapidly growing maturity of
superconducting circuits technology will play a major role in quantum
thermodynamics devices. In the context of the present setup, superconducting
technology is well suited for fabricating sets of qubits that interact
with their neighbors as well as with other baths \cite{PekolaSCengine}. 
\begin{acknowledgments}
Part of this work was supported by the COST Action MP1209 'Thermodynamics
in the quantum regime'. The support and warm hospitality of the Walmsley
group, and the physics department in Oxford University is greatly
acknowledged. The author cordially thanks the support the Kenneth
Lindsay trust fund. Finally, the author is in debt to Prof. Ronnie
Kosloff for his extraordinary support over the years.
\end{acknowledgments}

\bibliographystyle{apsrev4-1}
\bibliography{/Users/raam_uzdin/Dropbox/RaamCite}

\end{document}